\documentclass[aps,prl,twocolumn,amsmath,amssymb,graphicx,superscriptaddress]{revtex4-1}

\usepackage{bm}
\usepackage{subfigure}
\usepackage{graphicx}
\usepackage{amsthm}
\usepackage{notes2bib}
\usepackage{amsmath}
\usepackage{amssymb}
\usepackage{url}
\usepackage{enumerate}
\usepackage{epstopdf}
\usepackage{color}
\usepackage{mathtools}          
\usepackage{setspace}

\usepackage[usenames,dvipsnames,svgnames,table]{xcolor}
\usepackage{color}

\newcommand{\allblack}{\color{black}{}}

\newcommand{\mb}{\mathbf}
\newcommand{\mm}{\mathbb}
\newcommand{\torus}{{\mathbb{T}}}

\begin{document}
\title{Machine-learning inference of fluid variables from data using reservoir computing}
\author{Kengo Nakai}
\affiliation{Graduate School of Mathematical Sciences, The University of Tokyo, Tokyo 153-8914, Japan}
\author{Yoshitaka Saiki}
\affiliation{Graduate School of Business Administration, Hitotsubashi University, Tokyo 186-8601, Japan}
\affiliation{JST, PRESTO, Saitama 332-0012, Japan}
\affiliation{Institute for Physical Science and Technology, University of Maryland, College Park, Maryland 20742, USA}

\begin{abstract}
  We infer both microscopic and macroscopic behaviors of a 
  three-dimensional 
  chaotic fluid flow using reservoir computing.
  In our procedure of the inference, we assume no prior knowledge of a physical process of a fluid flow except that its behavior is complex but deterministic. 
  We present two ways of inference of the complex behavior; the first called partial-inference requires continued knowledge of partial time-series data during the inference as well as past time-series data,
while the second called full-inference requires only past time-series data as training data.
  For the first case, we are able to infer long-time motion of microscopic fluid variables. 
  For the second case, we show that the reservoir dynamics constructed from only past data of energy functions  
  can infer the future behavior of energy functions and reproduce the energy spectrum.
  It is also shown that we can infer a time-series data from only one measurement by using  the delay coordinates.
These implies that the obtained two reservoir systems constructed without the knowledge of microscopic data 
are equivalent to the dynamical systems describing macroscopic behavior of energy functions.
\end{abstract}
\date{\today}
\maketitle
\section{I. Introduction} 
Machine-learning has progressed significantly over the last decade
in various areas of physical sciences~\cite{Snyder_2012,Gabbard_2018, Mills_2018} 
after some theoretical works in the area of neural networks (See~\cite{Hornik_1989,Cybenko_1989} for examples.)\\
%
\indent
In fluid dynamics area~\textcite{Ling_2016} presents a method of using deep neural networks to learn a model for the Reynolds stress anisotropy tensor from high-fidelity simulation data (see also~\cite{Kutz_2018}). 
\textcite{Gamahara_2017} uses an artificial neural network to find a new subgrid model of the 
subgrid-scale stress in large-eddy simulation.
By using ``Long Short-Term Memory~(LSTM)''~\cite{Hochreiter_1997}, \textcite{Wan_2018} studies a data-assisted reduced-order modeling of extreme events in various dynamics including the Kolomogorov flow of the two-dimensional incompressible Navier--Stokes equation.
See also~\textcite{Vlachas_2018} for the result on the barotropic climate model.\\
%
\indent 
It is recently reported that reservoir computing, brain-inspired machine-learning framework that employs a data-driven dynamical system,
 is effective in the inference of a future such as time-series, frequency spectra 
and the Lyapunov spectra~\cite{Verstraeten_2007,Inubushi_2017, Zhixin_2017, Pathak_2017,Ibanez_2018,Pathak_2018,Antonik_2018}.
\textcite{Pathak_2017} exemplifies using the Lorenz system and the Kuramoto-Sivashinsky system that the model obtained by reservoir computing can generate an arbitrarily long time-series whose Lyapunov exponents approximate those of the input signal. \\
\indent A reservoir is a recurrent neural network whose internal parameters are not adjusted to fit the data in the training process. 
What is done is to train the reservoir by feeding it an input time-series and fitting a linear function of the reservoir state variables 
to a desired output time-series. 
Due to this approach of reservoir computing we can save a great amount of computational costs, 
which enables us to deal with a complex deterministic behavior. 
The framework was proposed as Echo-State Networks~\cite{Jaeger_2001,Jaeger_2004} and Liquid-State Machines~\cite{Maass_2002}.\\
\indent 
It is known that an inference of a fluid flow is difficult but important in both physical and industrial aspects. 
In this paper, we infer variables of a chaotic fluid flow by applying the method of reservoir computing without a prior knowledge of physical process. \\
\indent
After introducing the method of reservoir computing in Section II and a fluid flow in Section III, 
we explain how to apply the method to the inference of fluid variables, 
and show that inferences of both microscopic and macroscopic behaviors are successful in Sections IV and V, respectively. 
In Section VI, we exemplify that a time-series inference of high-dimensional dynamics is possible by using  delay coordinates,  even when  the number of measurements is smaller than the Lyapunov dimension of the attractor. 
Discussions and remarks are given in Section VII.
%
%
\section{II. Reservoir computing} 
Reservoir computing is recently used in the inference of complex dynamics~\cite{Zhixin_2017, Pathak_2017,Pathak_2018,Ibanez_2018,Lu_2018}. 
The reservoir computing focuses on the determination of a translation matrix from reservoir state variables to variables to be inferred 
(see eq.~(\ref{eq:output2})).
Here we review the outline of the method \cite{Jaeger_2004,Zhixin_2017}. We consider a dynamical system 
$$\frac{d\mb{\phi}}{dt}=\mb{f}(\mb{\phi}),$$
together with a pair of $\phi$-dependent, vector valued variables 
\begin{equation*}
\mb{u}=\mb{h}_1(\mb{\phi})\in \mm{R}^{M} 
~~\text{and}~~
\mb{s}=\mb{h}_2(\mb{\phi})\in \mm{R}^{P}.\label{eq:input}
\end{equation*} 
We seek a method for using the continued knowledge of 
$\mb{u}$ to determine an estimate of $\mb{s}$ as a function of time when direct measurement of $\mb{s}$ 
is not available, which we call the {\bf partial-inference}. 
We also consider the {\bf full-inference} for which we have a knowledge $\mb{u}$ only for $t\le T$.
Concerning the algorithm, this is just a variant of the partial-inference~\cite{Pathak_2017,Pathak_2018}, and will be explained later.\\
\indent 
The dynamics of the reservoir state vector 
$$\mb{r}\in \mm{R}^{N}~(N \gg M),$$
 is defined by
\begin{equation}
	\mb{r}(t+\Delta t)=(1-\alpha)\mb{r}(t)+\alpha \tanh(\mb{A}\mb{r}(t)+\mb{W}_{\text{in}}\mb{u}(t)
	),\label{eq:reservoir}
\end{equation}
where $\Delta t$ is a relatively short time step.
The matrix $\mb{A}$ is a weighted adjacency matrix of the reservoir layer, and the $M$-dimensional 
input $\mb{u}(t)$ is fed in to the $N$ reservoir nodes via a linear input weight matrix denoted by $\mb{W}_{\text{in}}$.
The parameter $\alpha$ ($0<\alpha\le 1$) in eq.~(\ref{eq:reservoir}) adjusts the nonlinearity of the dynamics of $\mb{r}$, 
and is chosen depending upon the complexity of the dynamics of measurements and the time step $\Delta t$. \\
\indent Each row of $\mb{W}_{\text{in}}$ has one nonzero element, chosen from a uniform distribution on $[-\sigma,\sigma]$.
The matrix $\mb{A}$ is chosen from a sparse random 
matrix in which the fraction of nonzero matrix elements is $(D_1+D_2)/N$, 
so that the average degree of a reservoir node is $D_1+D_2$. 
The $D_1$ non-zero components are chosen from a uniform distribution on $[-1, 1]$, and $D_2$ from that on $[-\gamma, \gamma]$ for $\gamma~(\ll 1)$, 
where $D_2$ non-zero components are introduced to reflect weak couplings among components of $\mb{r}$. 
Then we uniformly rescale all the elements of $\mb{A}$ so that the largest value of the magnitudes of its eigenvalues becomes $\rho$. \\
\indent The output, which is a $P$-dimensional vector, is taken to be a linear function of the reservoir state $\mb{r}$:
\begin{equation}
	\Hat{\mb{s}}(t)=\mb{W}_{out}\mb{r}(t)+\mb{c}.\label{eq:output}
\end{equation}
The reservoir state $\mb{r}$ evolves following eq.~(\ref{eq:reservoir}) with input 
$\mb{u}(t)$, 
starting from random initial state $\mb{r}(-\tau)$ whose elements are chosen from $(0, 1]$ in order not to diverge, 
where 
$\tau/\Delta t~(\gg 1)$  is the transient time.
We obtain $L=T/\Delta t$ steps of reservoir states $\{\mb{r}(l\Delta t)\}_{l=1}^{L}$ by eq.~(\ref{eq:reservoir}). 
Moreover, we record  the actual measurements of the state variables $\{\mb{s}(l\Delta t)\}_{l=1}^{L}$. \\
\indent We train the network by determining $\mb{W}_\text{out}$ and $\mb{c}$ 
so that the reservoir output approximates the measurement for $0< t \le T$~(training phase), which is the main part of this computation.
We do this by minimizing the following quadratic form with respect to $\mb{W}_\text{out}$ and $\mb{c}$:
\begin{equation}
\displaystyle\sum^{L}_{l=1} \|(\mb{W}_\text{out}\mb{r}(l\Delta t)+\mb{c})-\mb{s}(l\Delta t)\|^2
+\beta[Tr(\mb{W}_\text{out}\mb{W}^{T}_\text{out})],\label{eq:minimize}
\end{equation}
where $\|\mb{q}\|^2=\mb{q}^T \mb{q}$ for a vector $\mb{q}$,
and the second term is a regularization term introduced to avoid overfitting 
$\mb{W}_\text{out}$ for $\beta \ge 0$.
\allblack
When the training is successful, $\Hat{\mb{s}}(t)$ should approximate the desired unmeasured quantity $\mb{s}(t)$ for $t>T$~(inference phase). 
Following eq.~(\ref{eq:output}), we obtain 
\begin{equation}
	\Hat{\mb{s}}(t)=\mb{W}^*_\text{out}\mb{r}(t)+c^*,  \label{eq:output2}
\end{equation}
where $\mb{W}^*_\text{out}$ and $c^*$ denote the solutions for the minimizers 
of the quadratic form~(\ref{eq:minimize})~(see~\cite{Lukosevicius_2009}~P.140 for details):
\begin{align*}
	\mb{W}^*_\text{out}&=\delta\mb{S}\delta\mb{R}^{T}(\delta\mb{R}\delta\mb{R}^{T}+\beta\mb{I})^{-1},\\
	c^*&=-[\mb{W}^*_\text{out}\overline{\mb{r}}-\overline{\mb{s}}],
\end{align*}	
where $\overline{r}=\sum^{L}_{l=1} \mb{r}(l \Delta t)$/L, 
$\overline{s}=\sum^{L}_{l=1} \mb{s}(l \Delta t)/L$,
and $\mb{I}$ is the $N \times N$ identity matrix, $\delta\mb{R}$ (respectively, $\delta\mb{S}$) is the matrix 
whose $l$-th column is $\mb{r}(l\Delta t)-\overline{r}$ (respectively, $\mb{s}(l\Delta t)-\overline{s}$).\\
\begin{table*}[htb]
\normalsize
		\begin{tabular}{|l|l|r|r|r|} 
			\hline  	
           		  \multicolumn{2}{|c|}{parameter}& (a) & (b) & (c)\\ \hline
  			~$\tau$&transient time& 1000  &2500 &2350 \\ \hline
 			~$T$&training time&10000   &20000  &20000\\ \hline
  			~$M$&dimension of measurements&270  &9 &36  \\ \hline
  			~$P$&dimension of inferred variables &2 &9  &36 \\ \hline
	  		~$N$&number of reservoir nodes&6400 &3200 &3200\\ \hline
 	  		~$D_1$&parameter of determining elements of $\mb{A}$&60 &320 &120  \\ \hline
   			~$D_2$&parameter of determining elements of  $\mb{A}$&60 &0  &0\\ \hline
   			~$\gamma$&scale of input weights in $\mb{A}$ &0.1 &0  &0\\ \hline
	 	 	~$\rho$&maximal eigenvalue of $\mb{A}$ &1.0 &0.5 &0.5 \\  \hline
   			~$\sigma$&scale of input weights in $\mb{W}_{in}$&0.4 &0.3 &0.5 \\ \hline
	   		~$\alpha$&nonlinearity degree of reservoir dynamics&0.7 &0.3 &0.4  \\ \hline
   			~$\Delta t$&time step for reservoir dynamics &0.1 &0.25 &0.5 \\  \hline
 			~$\beta$&regularization parameter &0 &0.01  &0.1\\ \hline
 \end{tabular}
 		\caption{{\bf Sets of parameters for our reservoir computing.} The set (a) is used for the partial-inference of microscopic Fourier variables, whereas the set (b) is for the full-inference of macroscopic variables of energy functions and energy spectrum, 
 		and the set (c) is for the full-inference from only one measurement.
 		}
 		\label{tab:parameter}
\end{table*}
\indent 
In order to consider the effect of all the variables equally, we take the normalized value $\tilde{X}(t)$ for each variable $X(t)$, which will be used throughout the whole procedure of our reservoir computing:
  $$\tilde{X}(t)=[X(t)-X_1]/X_2,$$
where $X_1$ is the mean value and $X_2$ is the variance. 
When  we reconstruct $X(t)$ in the inference phase from $\tilde{X}(t)$, 
we employ $X_1$ and $X_2$ obtained in the training phase.
 Due to the normalization we can avoid adjustments of $\sigma$. \\
\begin{figure}[]
 \includegraphics[width=1.\columnwidth,height=0.705\columnwidth]{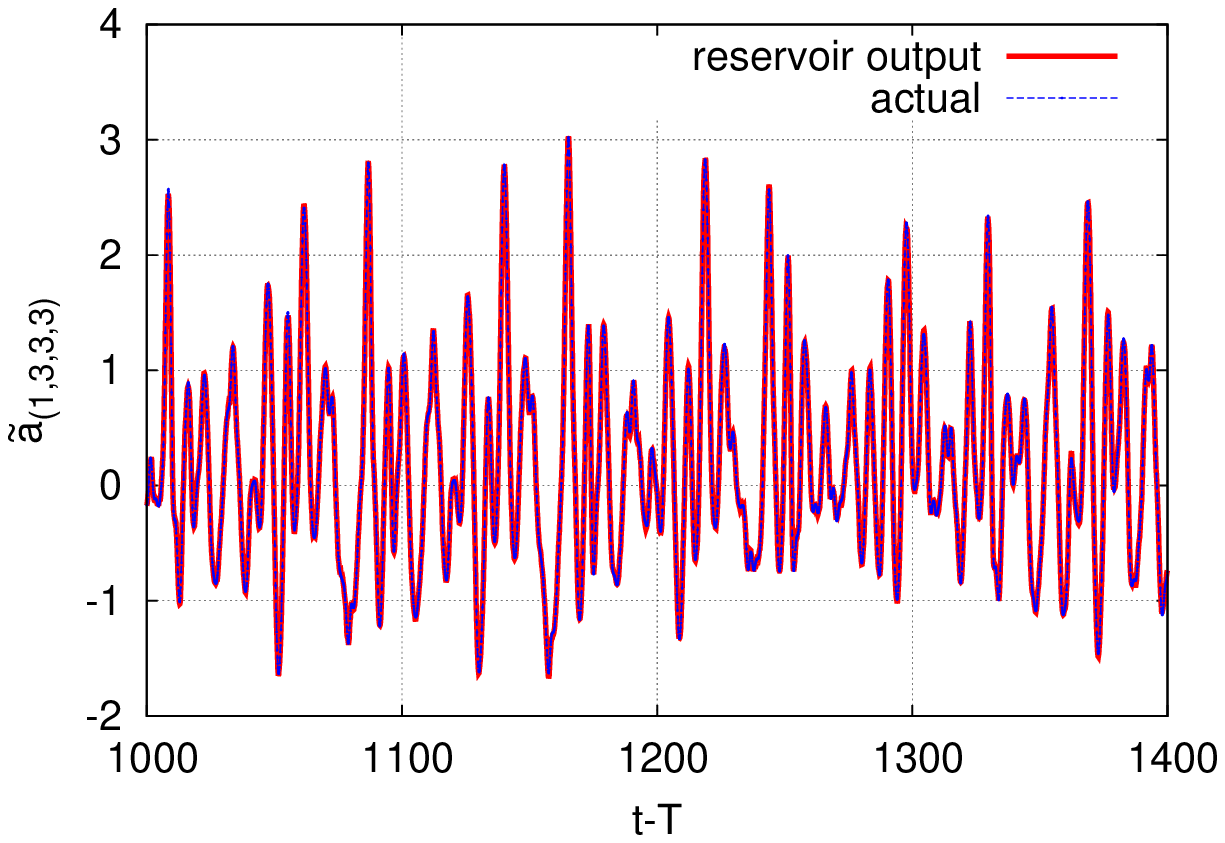}
  \includegraphics[width=1.\columnwidth,height=0.705\columnwidth]{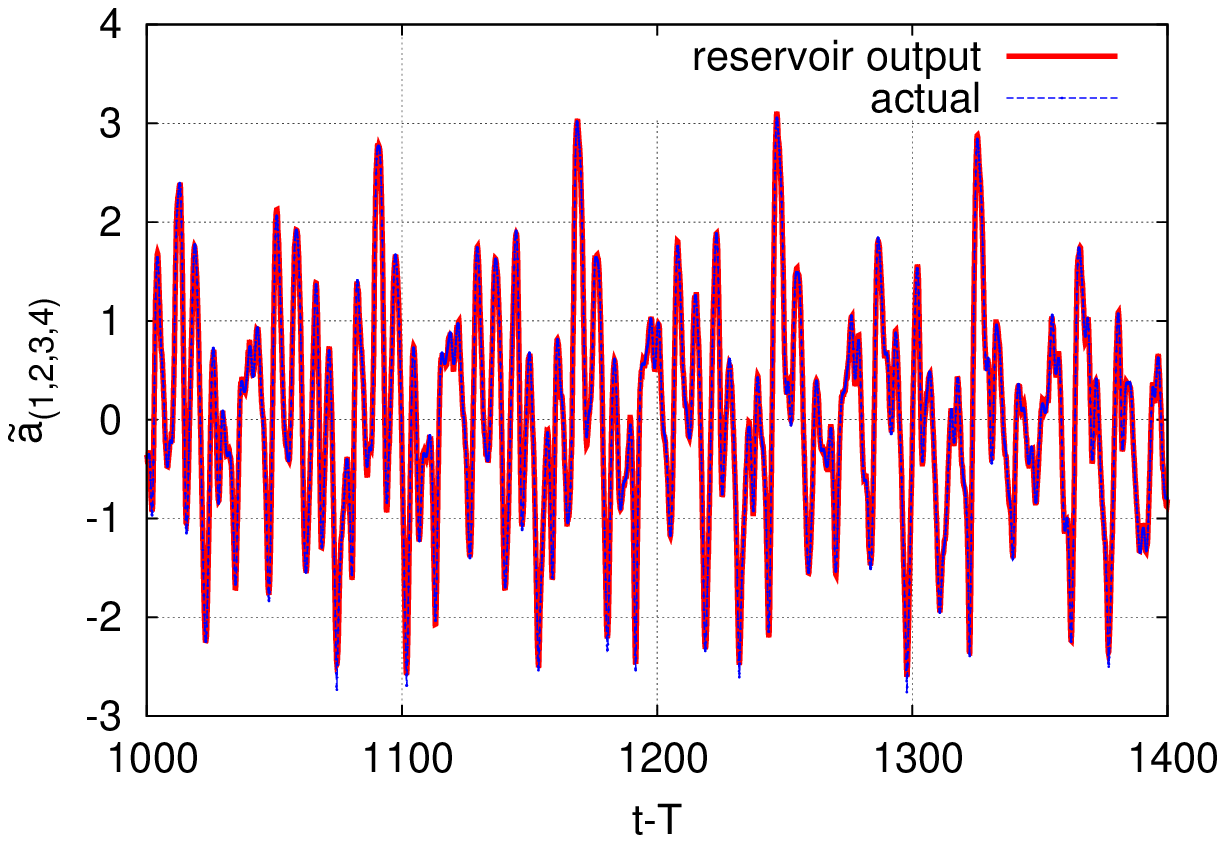}
\caption{ {\bf 
Partial-inference of time-series of microscopic variables in Fourier space of a fluid flow.}
Fourier variables ${\tilde{a}}_{\eta_1=(1, 3,3,3)}$~(top) and ${\tilde{a}}_{\eta_2=(1, 2,3,4)}$~(bottom) are inferred 
by using measured variables $\tilde{a}_{\eta}$ for $\eta \in S$ as well as the 
past time-series data for all the measured variables $\tilde{a}_{\eta}$ for $\eta \in S\cup\{\eta_1, \eta_2\}$.
We can observe that the inferred time-series almost coincide with the actual ones obtained by the direct numerical simulation of the Navier--Stokes equation even after sufficiently large time has passed since the training phase finished.
The inference errors in $l^1$-norm averaged over $t-T\in[0,2000]$ are $1.8 \%$ and $3.5 \%$ 
for $\tilde{a}_{\eta_1}$ and $\tilde{a}_{\eta_2}$, respectively.
}\label{fig:partial-micro}
\end{figure}
\section{III. Fluid flow} 
In order to generate measurements of the reservoir computing, 
we employ the direct numerical simulation of the incompressible three-dimensional Navier--Stokes equation 
under periodic boundary conditions:
	\begin{align*}
  \begin{cases}
		\partial_t v -\nu \Delta v
			+(v \cdot \nabla) v+\nabla \pi=f,~\nabla \cdot v=0, ~\mathbb{T}^3\times(0,\infty),\\
v\big| _{t=0}=v_0\quad \text{with $\nabla \cdot v_0=0$}, ~~~~~~~~~~~~~~~~~~~~\mathbb{T}^3,
  \end{cases}
	\end{align*}
where $\torus=[0,2\pi)$, $\nu>0$ is viscosity parameter, $\pi (x,t)$ is pressure, and $v(x,t)= (v_1(x,t),v_2(x,t),v_3(x,t))$ is velocity.
We use the Fourier spectral method~\cite{ishioka_1999} with $N_0(=9)$ modes in each direction, meaning that the system is approximated by 
$2(2 N_0+1)^3~(=13718)$-dimensional ordinary differential equations (ODEs).
The ODEs are integrated by the 4th-order Runge--Kutta method, and the forcing is input into the low-frequency variables at each time step so as to preserve the energy of the low-frequency part.
That is, both the real and the imaginary parts of the Fourier coefficient of the vorticity $\omega~(=\text{rot}~v)$, 
$$
	\mathcal{F}_{[\omega_{\zeta}]}(\kappa,t):= \dfrac{1}{(2\pi)^3} \displaystyle\int_{\mathbb{T}^3}
			\omega_{\zeta}(x, t)
		e^{-i(\kappa\cdot x)}dx, 
$$
 are kept constant  for $\zeta=1,2$, $\kappa = (1,0,0), (0,1,0)$. 
 We use  an initial condition, which  has energy only in the low-frequency variables. 
See~\cite{ishioka_1999} for the details.  
\section{IV. Partial-inference of microscopic variables: Fourier variables of velocity.}
We consider the absolute value of Fourier variables of velocity $\mathcal{F}_{[v_{\zeta}]}(\kappa,t)$ as the representative microscopic variables:
	\begin{align}
		a_{\eta} (t)= \left| \mathcal{F}_{[v_{\zeta}]}(\kappa,t) \right|
		:= \left| \dfrac{1}{(2\pi)^3} \int_{\mathbb{T}^3}
			v_{\zeta}(x, t)
		e^{-i(\kappa\cdot x)}dx  \right|, \label{eq:fourier-coefficient}
	\end{align}
	where 
		$\eta = (\zeta, \kappa)\in S_0:= \{(\zeta, \kappa_1, \kappa_2, \kappa_3) \in \mathbb{Z}^4 |~ \zeta\in \{1,2,3\}, \kappa_1, \kappa_2, \kappa_3 \in [-N_0,N_0] \}$.
Since $v$ is real, $a_{(\cdot, \kappa_1, \kappa_2, \kappa_3)}=a_{(\cdot, -\kappa_1, -\kappa_2, -\kappa_3)}$. 
The reason why we take the absolute value in eq.~(\ref{eq:fourier-coefficient}) is to kill the rotational invariance of a complex variable and to make an inference possible.
We  choose a chaotic parameter $\nu=0.05862$, 
and set $\mb{u}(t)$ as the time-series of $M=270$ Fourier variables $\tilde{a}_{\eta}$, 
where $\eta \in S:= \{(2, \pm  \kappa_1, \kappa_2, \kappa_3) \in \mathbb{Z}^4 |~1 \leq \kappa_1 \leq N_0, \kappa_1 \leq \kappa_2 \leq \kappa_3 \leq \kappa_1+4 \}$ and each component is taken $\bmod~N_0$, 
 that is, 
	\begin{align*}
	\mb{u}(t)&=(\{ \tilde{a}_{\eta} \}_{\eta \in S})^{t}.
	\end{align*}
We also set 
	\begin{align*}
	\mb{s}(t)&=(\tilde{a}_{(1, 3,3,3)}, \tilde{a}_{(1, 2,3,4)})^{t},   
	\end{align*}
where $(1, 3,3,3),  (1, 2,3,4) \notin S$. 
Under the set of parameters in TABLE \ref{tab:parameter}~(a)
we infer the time-series $\mb{s}(t)$,  
which is successful for quite a long time (see Fig.~\ref{fig:partial-micro}).\\
\indent 
The choice of variables to be trained is not very significant in this study, because the attractor does not show a homogeneous 
isotropic turbulence, and it has less symmetries.
We can see from the Poincar\'e section of the microscopic variables that the flow is not isotropic and 
indeterminacy in inference due to the continuous symmetry does not appear. 
However, by training variables with different types of behaviors, we can construct a reservoir model 
in less computational costs with lower dimension $N$ of the reservoir system. 
In fact, we confirmed that we can infer some other fluid variables including both low-frequency and high-frequency variables 
from some other training variables.
We found that an inference of a high-frequency variable tends to be more difficult, maybe because of the stronger intermittency.
Remark that $D_2$ is useful to represent non-local relatively weak interactions among microscopic variables in the partial inference. 
\begin{figure}
\includegraphics[width=1.0\columnwidth,height=0.705\columnwidth]{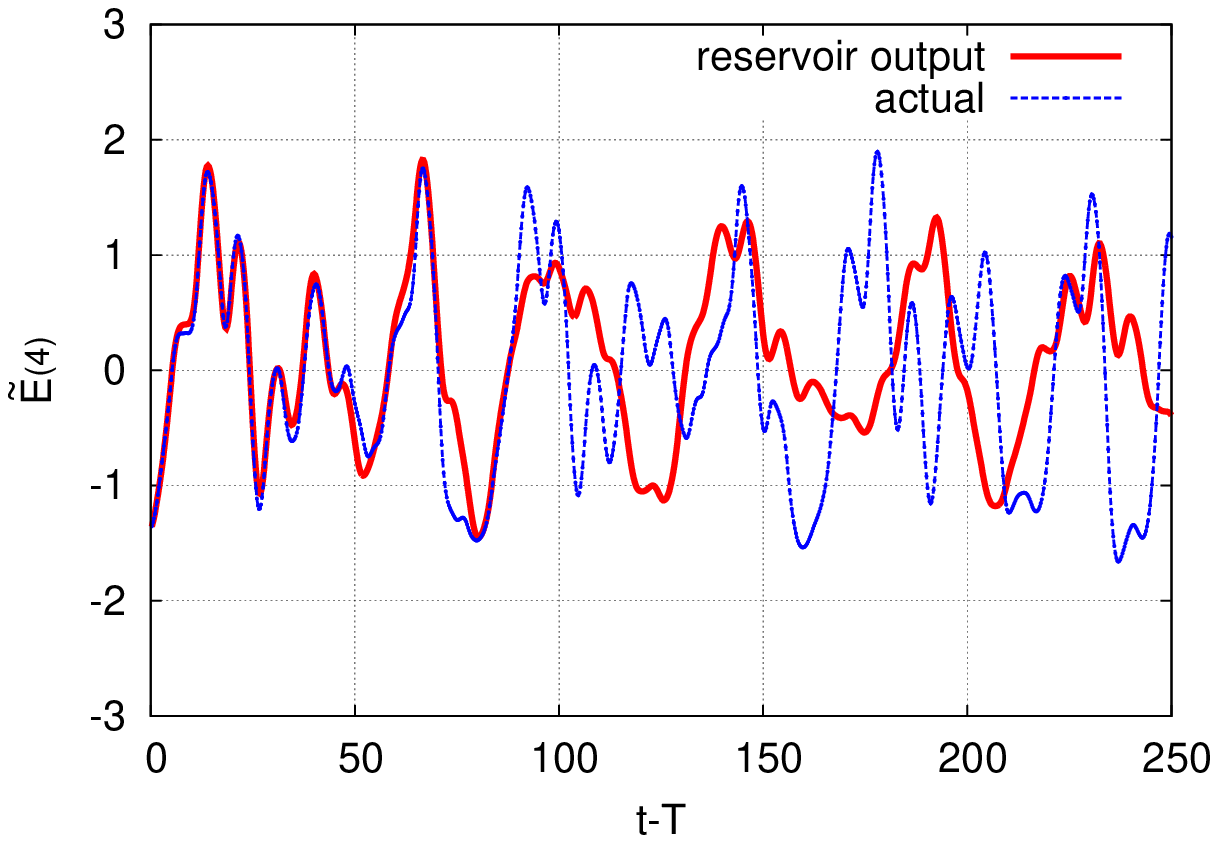}
\includegraphics[width=1.0\columnwidth,height=0.705\columnwidth]{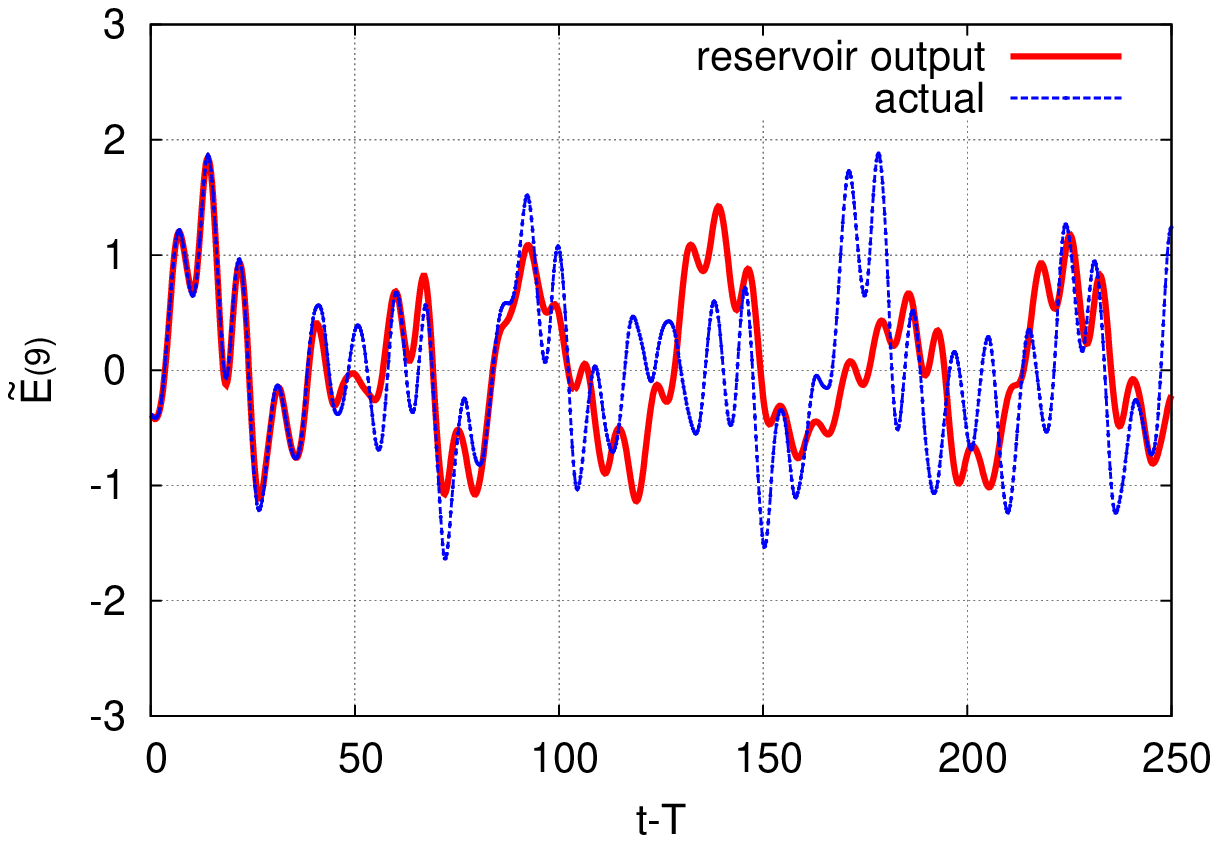}
\includegraphics[width=1.0\columnwidth,height=0.705\columnwidth]{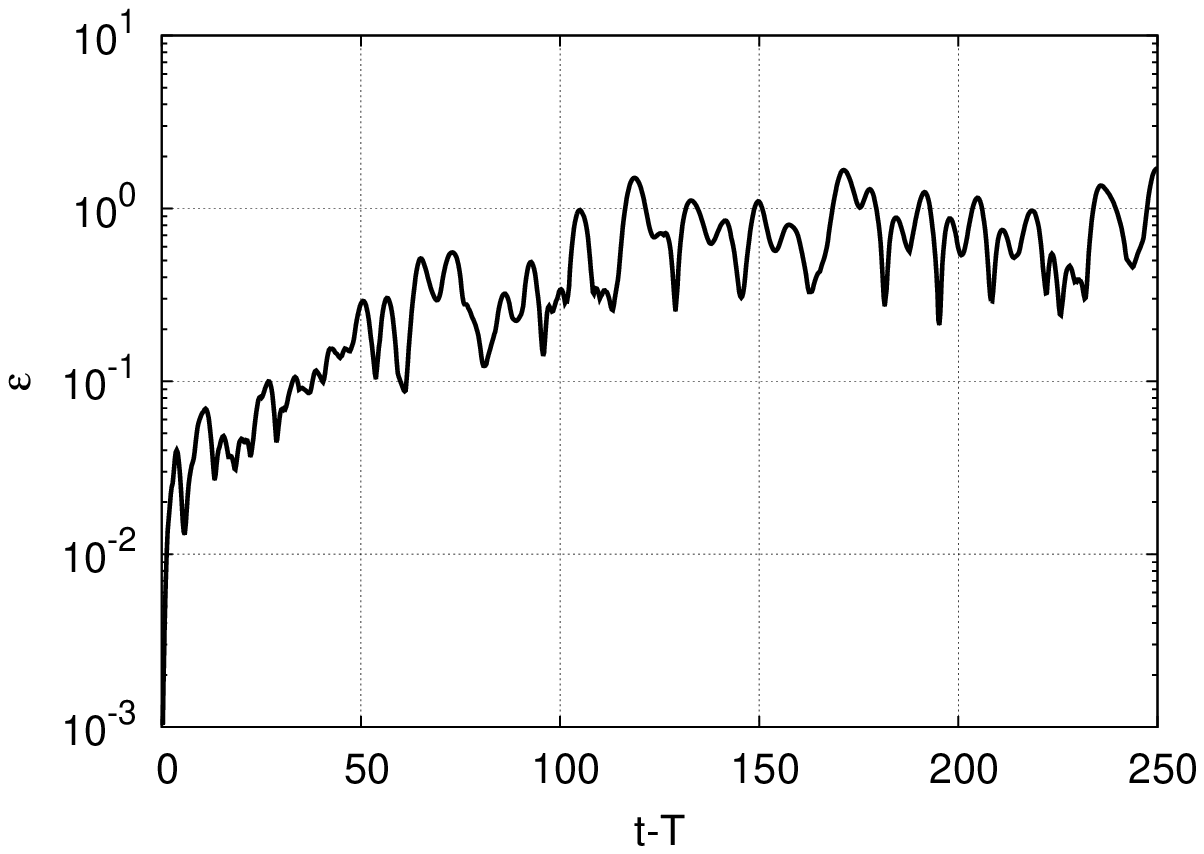}
\caption{ {\bf 
Full-inference of time-series of macroscopic variables of a fluid flow.}
Time-series of energy function ${\tilde{E}}(k,t)$ for $k=4$ (top) and $9$ (middle)  are inferred from the reservoir system 
in comparison with that of a reference data obtained by the direct numerical simulation of the Navier--Stokes equation.
The inference error defined by $\varepsilon(t)=\sum^{N_0}_{k=1}|\tilde{E}(k,t)-\hat{\tilde{E}}(k,t)|/N_0$ $(N_0=9)$ is shown to grow exponentially with time up to $t-T=100$ (bottom), which is inevitable for a chaotic behavior of a fluid flow.
The growth of error within a short time highly depends on the direction of the perturbation vector
 $\{\tilde{E}(\cdot,T+\Delta t)-\hat{\tilde{E}}(\cdot,T+\Delta t)\}$, 
 \allblack
 and its slope can vary in different settings.
}\label{fig:full-macro}
\end{figure}
\begin{figure}
   \includegraphics[width=1.0\columnwidth, height=0.705\columnwidth]{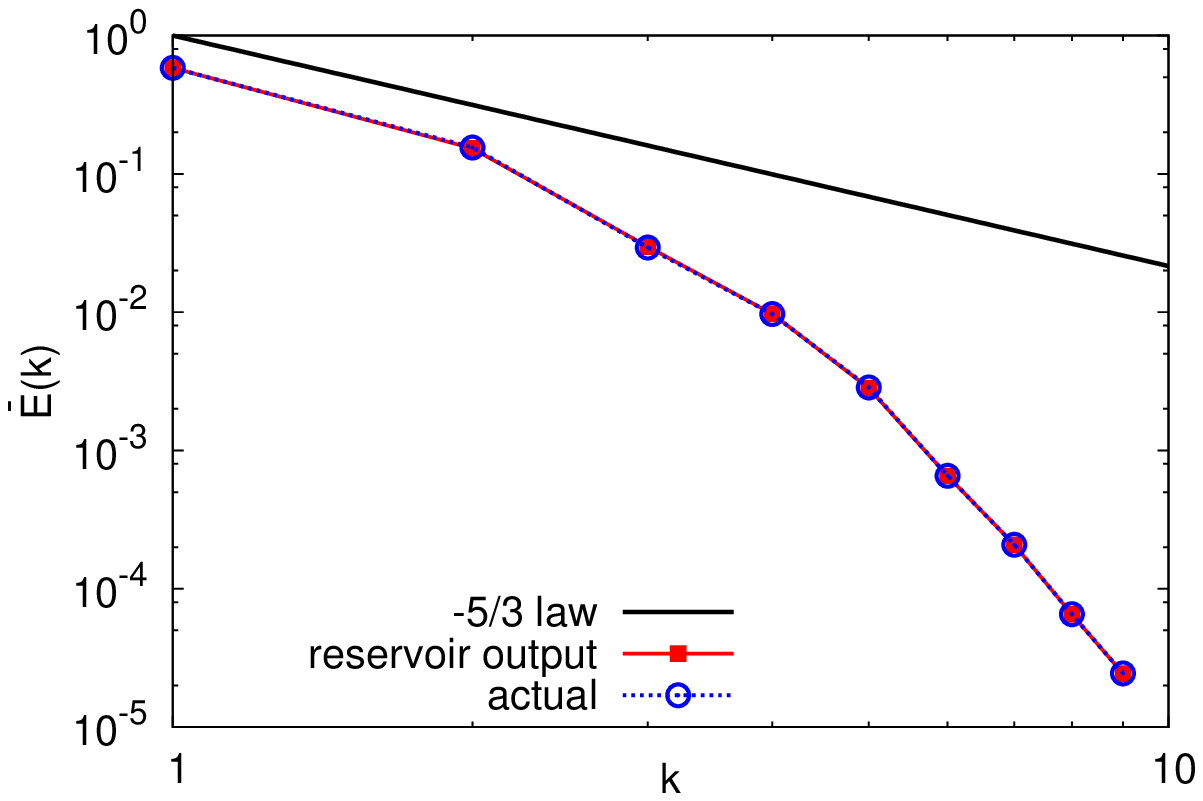}	
\caption{
{\bf 
Energy spectrum $\overline{E}(k)$ reproduced from the reservoir computing.}
The spectrum is obtained from the full-inference of an energy function $E(k,t)$, which is compared with that for a reference data obtained by the direct numerical simulation of the Navier--Stokes equation.
The coincidence of the two energy spectra implies that the reservoir system captures the dynamics of a fluid flow in statistical sense, 
even after the time-series inference has failed due to the chaotic property (see Fig.~\ref{fig:full-macro}). 
The Kolmogorov $-5/3$ law of the energy spectrum is shown as a reference.
The relative error of inferred variable $\overline{\hat{E}}(k)$ from $\overline{E}(k)$ $(k=1,\cdots,9)$ is up to $1.3\%$. 
}\label{fig:full-macro-average}
\end{figure}
\section{V. Full-inference of macroscopic variables: Energy function and Energy spectrum} 
We study an energy function as the representative of a macroscopic variable. 
We set $\nu=0.058$ for which the flow is more turbulent than the previous case.
However, the complexity of the dynamics is much less 
than that for a microscopic variable for the same viscosity.  
This is because the energy function can be thought of as an averaged quantity of many microscopic variables. 
The energy function $E_0(k, t)$ for wavenumber $k \in \mathbb{N}$ is defined by 
	\begin{align*}
		E_0(k, t):= \dfrac{1}{2} \int_{D_k} {\sum_{\zeta=1}^{3}}
			\left| 
				\mathcal{F}_{[v_{\zeta}]}(\kappa, t)
			\right|^2  d\kappa, 
	\end{align*}
where $D_k:=\{ \kappa \in \mathbb{Z}^3| k-0.5 \leq |\kappa| < k+0.5 \}$. 
See eq.~(\ref{eq:fourier-coefficient}) for the expression of $\mathcal{F}_{[v_{\zeta}]}(\kappa, t)$.
In order to get rid of the high-frequency fluctuation, we take the
short-time average 
$$E(k,t)=\sum_{s=t-99\Delta s}^{t}E_0(k,s)/100,$$ 
where $\Delta s=0.05$ is the time step of the integration of the Navier--Stokes equation.  
This helps us to obtain essential low-frequency dynamics of an energy function and infer its time-series 
with less computational costs with lower dimension $N$ of the reservoir vectors. 
The averaged energy function $E(k,t)$ will be called an energy function hereafter.\\
%
\indent In the training phase for $t \in (0,T]$, $\mb{W}^{*}_\text{out}$ and $c^{*}$ are determined by setting
	\begin{align*}
	\mb{u}(t)&=(\tilde{E}(1,t), \tilde{E}(2,t),\cdots,\tilde{E}(9,t))^{t},  \\
	\mb{s}(t)&=(\tilde{E}(1,t), \tilde{E}(2,t),\cdots,\tilde{E}(9,t))^{t}, 
	\end{align*}
and by following the same procedure as the partial-inference.
In the inference phase for $t>T$, eq.(\ref{eq:reservoir}) is written as 
\begin{equation*}
\mb{r}(t+\Delta t)=(1-\alpha)\mb{r}(t)+\alpha \tanh(\mb{A}\mb{r}(t)+\mb{W}_{\text{in}}\Hat{\mb{s}}(t)),\label{eq:full-reservoir}
\end{equation*}
by setting $\mb{u}(t)$ as 
	\begin{align*}
	\Hat{\mb{s}}(t)=(\hat{\tilde{E}}(1,t), \hat{\tilde{E}}(2,t),\cdots,\hat{\tilde{E}}(9,t))^{t} 
	\end{align*}
	 obtained from eq.~(\ref{eq:output2}).
A set of parameters employed here is shown in TABLE \ref{tab:parameter}~(b). \\
\indent We found that an inference of energy functions is successful 
for some time after finishing training $9$-dimensional time-series data of energy functions. 
The two cases for $\tilde{E}(4,t)$ and $\tilde{E}(9,t)$ are shown in Fig.~\ref{fig:full-macro}~(top)(middle). 
The failure in the long-term time-series inference is inevitable just due to the sensitive dependence on initial condition of a chaotic property of the fluid flow. 
In fact, the growth rate of error in the energy functions is shown to be exponential for $t-T\lesssim 100$ in Fig.~\ref{fig:full-macro}~(bottom).
However, the energy spectrum $\overline{E}(k)=\langle E(k,t) \rangle$, the time average of an energy function $E(k,t)$,  can be reproduced from the inferred time-series data for $1000<t-T<2000$ (Fig.~\ref{fig:full-macro-average}).
This implies that the reservoir system constructed without the knowledge of microscopic variables captures statistical property correctly,
and that the obtained system can be understood as a chaotic dynamical system describing a behavior of energy functions.\\
\indent
\section{VI. Full-inference of Macroscopic variable from only one measurement using delay coordinates} 
\begin{figure}[]
 \includegraphics[width=1.\columnwidth,height=0.705\columnwidth]{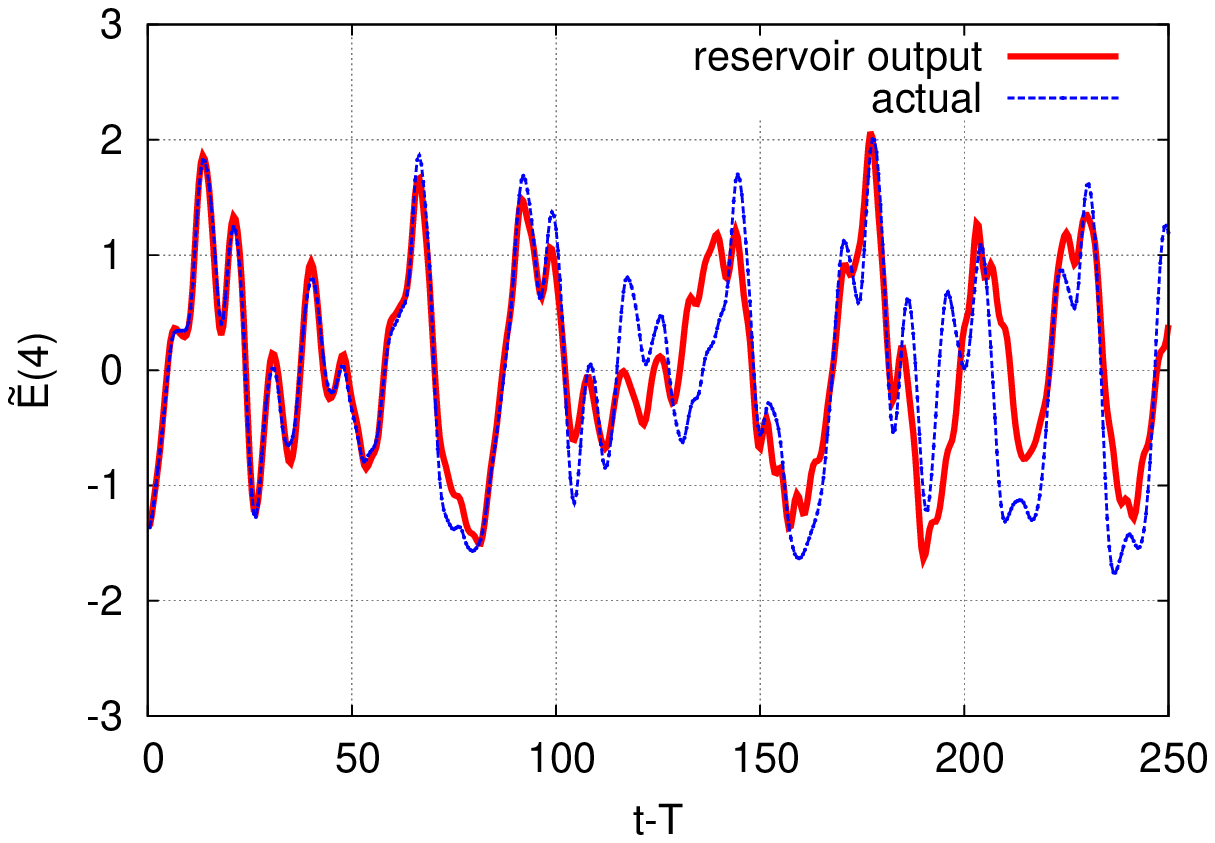} 
  \includegraphics[width=1.\columnwidth,height=0.705\columnwidth]{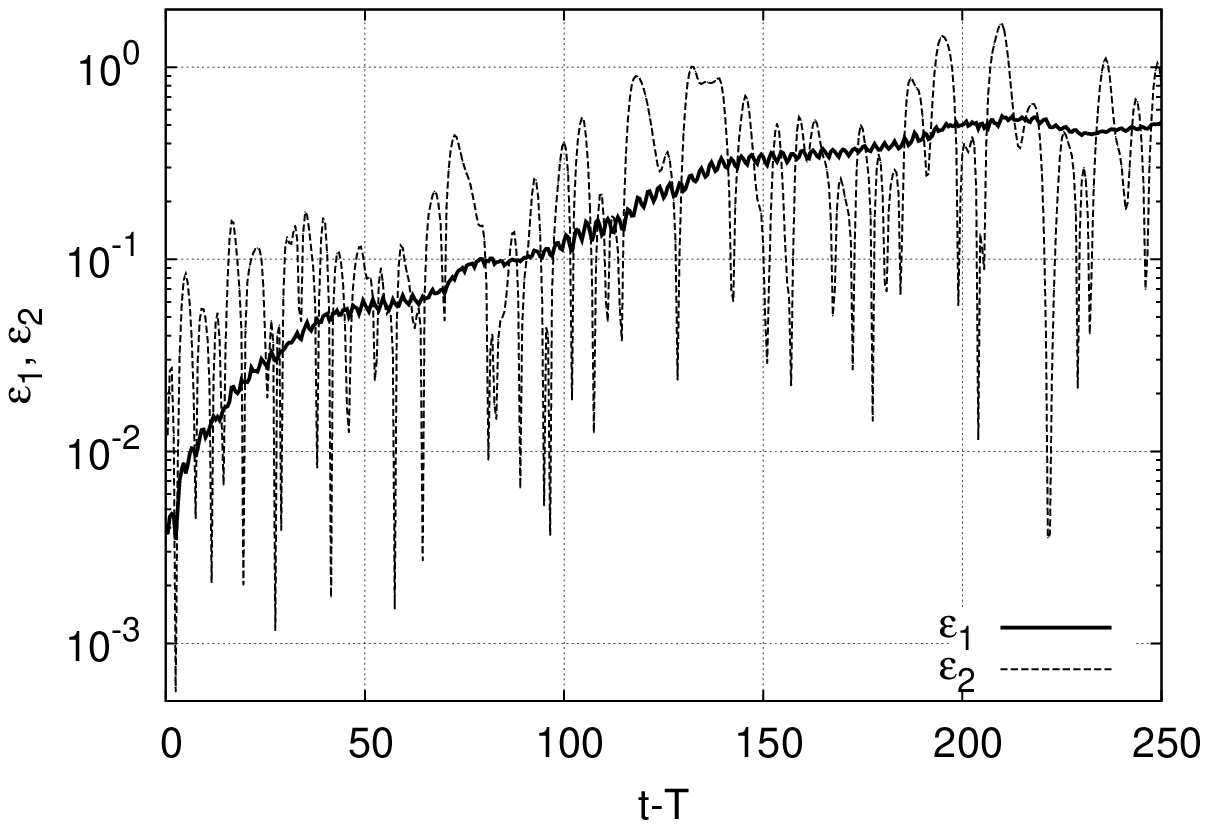} 
\caption{ 
{\bf 
Full-inference of a macroscopic variable using the delay coordinates of only one measurement.
}
We infer an energy function ${\tilde{E}}(4,t)$ for the same time range as in Fig.~\ref{fig:full-macro}~(top)
from only one measurement ${\tilde{E}}(4,t)$. 
The inferred time-series of  ${\tilde{E}}(4,t)$ is shown together with a reference data obtained 
by the direct numerical simulation of the Navier--Stokes equation~(top).  
Errors for the inference $\varepsilon_1(t)=\sum^{35}_{n=0}|\tilde{E}(4,t-n \Delta \tau)-\hat{\tilde{E}}(4,t-n \Delta \tau)|^2/36$  and $\varepsilon_2(t)=|\tilde{E}(4,t)-\hat{\tilde{E}}(4,t)|$ 
are shown~(bottom).
}\label{fig:energy_function_delay}
\end{figure}

In various experiments and observations of high-dimensional complex phenomena, 
there are usually much smaller number of measurements than the Lyapunov dimensions of the attractor.
Even in such cases we can infer a time-series data by generating high-dimensional input data $\mb{u}$ for the reservoir computation through the delay-coordinate embedding method~\cite{Takens,Embedology}. \\
%
\indent
Here we exemplify a full-inference of an energy function $E(4,t)$ for the same flow as in Section V, 
by assuming that the accessible measurement is limited to only one variable $E(4,t)$ 
among $9$ measurements $E(k,t)~(k=1,\cdots,9)$ used in Section V.
In order to overcome  the lack of sufficiently large number of  measurements,
we introduce $36$-dimensional delay-coordinate function  with a time delay  $\Delta\tau=2.5$, that is, 
\begin{align*}
\mb{u}(t)&=(\tilde{E}(4,t), \tilde{E}(4,t-\Delta\tau),\cdots,\tilde{E}(4,t-35\Delta\tau))^{t}, \\
\mb{s}(t)&=(\tilde{E}(4,t), \tilde{E}(4,t-\Delta\tau),\cdots,\tilde{E}(4,t-35\Delta\tau))^{t} .
\end{align*}
An inferred time-series of $\tilde{E}(4,t)$ is shown in Fig.~\ref{fig:energy_function_delay}, which is as successful 
as the case when there are $9$ measurements in Fig.~\ref{fig:full-macro}~(top). 
A set of parameters employed here is shown in TABLE \ref{tab:parameter}~(c). \\
%
%
%
\section{VII. Discussion and remarks}
We have succeeded in inferring time-series of both microscopic and macroscopic variables of a 
three-dimensional 
fluid flow by machine-learning technique using reservoir computing.  
The method is especially useful in generating an arbitrarily long time-series data of macroscopic variables as well as a statistical property 
 with small computational costs. 
 That is, in order to generate a time-series data of a macroscopic variable of a fluid flow, 
we do not need to refer microscopic behaviors. 
It takes roughly $1/80$ of time to obtain a time-series of the energy functions $E(k)$ with the same time-lengths, when we use the model constructed by the reservoir computation. The Navier--Stokes equation is calculated by 13718-dimensional ODEs with the $4$-stage Runge--Kutta method (time step $0.05$), whereas the model is calculated by 3200-dimensional map whose iterate corresponds to the time step $0.25$. \\
\indent The difficulty in the construction of a reservoir model can vary mainly depending on the viscosity $\nu$. 
As the degree of turbulence increases by decreasing $\nu$,  longer training time $T$ and higher dimension $N$ of the reservoir state vector $\mb{r}\in \mm{R}^{N}$ are required. However, for macroscopic variables the construction is relatively easy, even when the flow is turbulent.
Because the degree of instability of a macroscopic behavior is relatively low in comparison with that of a microscopic behavior. \\
\indent It is expected that our procedure will work, even if a high-frequency noise is added to the training data, because even in our current computation we have applied a low-pass filter for the inference of macroscopic variables.
Although our approach focuses on constructing a model for a fluid flow with a fixed parameter $\nu$, it will be very interesting to consider a framework of the construction of a model with a parameter. \\
\indent When we do numerical computation of the Navier--Stokes equation, we employ some discretized expressions using 
Fourier spectrum method, finite difference method and finite element method. 
The obtained reservoir system constructed from data can be understood as one of such expressions, describing a macroscopic (or a microscopic) dynamics of a  fluid flow.\\
\indent It is known that there is a difficulty in obtaining a closed form equation of macroscopic behavior of a fluid flow 
from the Navier--Stokes equation analytically, so called a ``closure problem''. That is, in order to express the dynamics of the $n$-th moment variables, the $n+1$-th moment variables are required for any positive integer $n$.
Our study on the data-driven modeling may give us insights on this kind of problem.
For a relatively large value of $\nu$ considered in our paper, $\{E(k)\}_{k=1}^{K}$ seems to be enough for
representing the dynamics of $E(k)$, whereas $\{E(k)\}_{k=1}^{K}$ will not be enough 
for more turbulent case with a smaller value of $\nu$, even if $K$ is chosen large enough. 
In such a case time-delay variables can be used for generating high-dimensional input data as are used in Section VI. 
\section{Acknowledgements} 
We would like to thank anonymous referees for the critical reading of the manuscript and giving us some insightful comments to improve our paper.
KN was supported by the Leading Graduate Course for Frontiers of Mathematical Sciences and Physics (FMSP) at the University of Tokyo.
YS was supported by the JSPS KAKENHI Grant No.17K05360 and JST PRESTO JPMJPR16E5.
Part of the computation was supported by the Collaborative Research Program for Young $\cdot$ Women Scientists of ACCMS and IIMC, Kyoto University.

\bibliography{reservoir_bibliography,qr_bibliography}

\begin{thebibliography}{26}%
\makeatletter
\providecommand \@ifxundefined [1]{%
 \@ifx{#1\undefined}
}%
\providecommand \@ifnum [1]{%
 \ifnum #1\expandafter \@firstoftwo
 \else \expandafter \@secondoftwo
 \fi
}%
\providecommand \@ifx [1]{%
 \ifx #1\expandafter \@firstoftwo
 \else \expandafter \@secondoftwo
 \fi
}%
\providecommand \natexlab [1]{#1}%
\providecommand \enquote  [1]{``#1''}%
\providecommand \bibnamefont  [1]{#1}%
\providecommand \bibfnamefont [1]{#1}%
\providecommand \citenamefont [1]{#1}%
\providecommand \href@noop [0]{\@secondoftwo}%
\providecommand \href [0]{\begingroup \@sanitize@url \@href}%
\providecommand \@href[1]{\@@startlink{#1}\@@href}%
\providecommand \@@href[1]{\endgroup#1\@@endlink}%
\providecommand \@sanitize@url [0]{\catcode `\\12\catcode `\$12\catcode
  `\&12\catcode `\#12\catcode `\^12\catcode `\_12\catcode `\%12\relax}%
\providecommand \@@startlink[1]{}%
\providecommand \@@endlink[0]{}%
\providecommand \url  [0]{\begingroup\@sanitize@url \@url }%
\providecommand \@url [1]{\endgroup\@href {#1}{\urlprefix }}%
\providecommand \urlprefix  [0]{URL }%
\providecommand \Eprint [0]{\href }%
\providecommand \doibase [0]{http://dx.doi.org/}%
\providecommand \selectlanguage [0]{\@gobble}%
\providecommand \bibinfo  [0]{\@secondoftwo}%
\providecommand \bibfield  [0]{\@secondoftwo}%
\providecommand \translation [1]{[#1]}%
\providecommand \BibitemOpen [0]{}%
\providecommand \bibitemStop [0]{}%
\providecommand \bibitemNoStop [0]{.\EOS\space}%
\providecommand \EOS [0]{\spacefactor3000\relax}%
\providecommand \BibitemShut  [1]{\csname bibitem#1\endcsname}%
\let\auto@bib@innerbib\@empty
\bibitem [{\citenamefont {Snyder}\ \emph {et~al.}(2012)\citenamefont {Snyder},
  \citenamefont {Rupp}, \citenamefont {Hansen}, \citenamefont {M{\"u}ller},\
  and\ \citenamefont {Burke}}]{Snyder_2012}%
  \BibitemOpen
  \bibfield  {author} {\bibinfo {author} {\bibfnamefont {J.~C.}\ \bibnamefont
  {Snyder}}, \bibinfo {author} {\bibfnamefont {M.}~\bibnamefont {Rupp}},
  \bibinfo {author} {\bibfnamefont {K.}~\bibnamefont {Hansen}}, \bibinfo
  {author} {\bibfnamefont {K.-R.}\ \bibnamefont {M{\"u}ller}}, \ and\ \bibinfo
  {author} {\bibfnamefont {K.}~\bibnamefont {Burke}},\ }\href@noop {}
  {\bibfield  {journal} {\bibinfo  {journal} {Phys. Rev. Lett.}\ }\textbf
  {\bibinfo {volume} {108}},\ \bibinfo {pages} {253002} (\bibinfo {year}
  {2012})}\BibitemShut {NoStop}%
\bibitem [{\citenamefont {Gabbard}\ \emph {et~al.}(2018)\citenamefont
  {Gabbard}, \citenamefont {Williams}, \citenamefont {Hayes},\ and\
  \citenamefont {Messenger}}]{Gabbard_2018}%
  \BibitemOpen
  \bibfield  {author} {\bibinfo {author} {\bibfnamefont {H.}~\bibnamefont
  {Gabbard}}, \bibinfo {author} {\bibfnamefont {M.}~\bibnamefont {Williams}},
  \bibinfo {author} {\bibfnamefont {F.}~\bibnamefont {Hayes}}, \ and\ \bibinfo
  {author} {\bibfnamefont {C.}~\bibnamefont {Messenger}},\ }\href@noop {}
  {\bibfield  {journal} {\bibinfo  {journal} {Phys. Rev. Lett.}\ }\textbf
  {\bibinfo {volume} {120}},\ \bibinfo {pages} {141103} (\bibinfo {year}
  {2018})}\BibitemShut {NoStop}%
\bibitem [{\citenamefont {Mills}\ and\ \citenamefont
  {Tamblyn}(2018)}]{Mills_2018}%
  \BibitemOpen
  \bibfield  {author} {\bibinfo {author} {\bibfnamefont {K.}~\bibnamefont
  {Mills}}\ and\ \bibinfo {author} {\bibfnamefont {I.}~\bibnamefont
  {Tamblyn}},\ }\href@noop {} {\bibfield  {journal} {\bibinfo  {journal} {Phys.
  Rev. E}\ }\textbf {\bibinfo {volume} {97}},\ \bibinfo {pages} {032119}
  (\bibinfo {year} {2018})}\BibitemShut {NoStop}%
\bibitem [{\citenamefont {Hornik}\ \emph {et~al.}(1989)\citenamefont {Hornik},
  \citenamefont {Stinchcombe},\ and\ \citenamefont {White}}]{Hornik_1989}%
  \BibitemOpen
  \bibfield  {author} {\bibinfo {author} {\bibfnamefont {K.}~\bibnamefont
  {Hornik}}, \bibinfo {author} {\bibfnamefont {M.}~\bibnamefont {Stinchcombe}},
  \ and\ \bibinfo {author} {\bibfnamefont {H.}~\bibnamefont {White}},\
  }\href@noop {} {\bibfield  {journal} {\bibinfo  {journal} {Neural Networks}\
  }\textbf {\bibinfo {volume} {2}},\ \bibinfo {pages} {359} (\bibinfo {year}
  {1989})}\BibitemShut {NoStop}%
\bibitem [{\citenamefont {Cybenko}(1989)}]{Cybenko_1989}%
  \BibitemOpen
  \bibfield  {author} {\bibinfo {author} {\bibfnamefont {G.}~\bibnamefont
  {Cybenko}},\ }\href@noop {} {\bibfield  {journal} {\bibinfo  {journal}
  {Mathematics of control, signals and systems}\ }\textbf {\bibinfo {volume}
  {2}},\ \bibinfo {pages} {303} (\bibinfo {year} {1989})}\BibitemShut {NoStop}%
\bibitem [{\citenamefont {Ling}\ \emph {et~al.}(2016)\citenamefont {Ling},
  \citenamefont {Kurzawski},\ and\ \citenamefont {Templeton}}]{Ling_2016}%
  \BibitemOpen
  \bibfield  {author} {\bibinfo {author} {\bibfnamefont {J.}~\bibnamefont
  {Ling}}, \bibinfo {author} {\bibfnamefont {A.}~\bibnamefont {Kurzawski}}, \
  and\ \bibinfo {author} {\bibfnamefont {J.}~\bibnamefont {Templeton}},\
  }\href@noop {} {\bibfield  {journal} {\bibinfo  {journal} {J. Fluid Mech.}\
  }\textbf {\bibinfo {volume} {807}},\ \bibinfo {pages} {155} (\bibinfo {year}
  {2016})}\BibitemShut {NoStop}%
\bibitem [{\citenamefont {Kutz}(2017)}]{Kutz_2018}%
  \BibitemOpen
  \bibfield  {author} {\bibinfo {author} {\bibfnamefont {J.~N.}\ \bibnamefont
  {Kutz}},\ }\href@noop {} {\bibfield  {journal} {\bibinfo  {journal} {J. Fluid
  Mech.}\ }\textbf {\bibinfo {volume} {814}},\ \bibinfo {pages} {1} (\bibinfo
  {year} {2017})}\BibitemShut {NoStop}%
\bibitem [{\citenamefont {Gamahara}\ and\ \citenamefont
  {Hattori}(2017)}]{Gamahara_2017}%
  \BibitemOpen
  \bibfield  {author} {\bibinfo {author} {\bibfnamefont {M.}~\bibnamefont
  {Gamahara}}\ and\ \bibinfo {author} {\bibfnamefont {Y.}~\bibnamefont
  {Hattori}},\ }\href@noop {} {\bibfield  {journal} {\bibinfo  {journal} {Phys.
  Rev. Fluids}\ }\textbf {\bibinfo {volume} {2}},\ \bibinfo {pages} {054604}
  (\bibinfo {year} {2017})}\BibitemShut {NoStop}%
\bibitem [{\citenamefont {Hochreiter}\ and\ \citenamefont
  {Schmidhuber}(1997)}]{Hochreiter_1997}%
  \BibitemOpen
  \bibfield  {author} {\bibinfo {author} {\bibfnamefont {S.}~\bibnamefont
  {Hochreiter}}\ and\ \bibinfo {author} {\bibfnamefont {J.}~\bibnamefont
  {Schmidhuber}},\ }\href@noop {} {\bibfield  {journal} {\bibinfo  {journal}
  {Neural computation}\ }\textbf {\bibinfo {volume} {9}},\ \bibinfo {pages}
  {1735} (\bibinfo {year} {1997})}\BibitemShut {NoStop}%
\bibitem [{\citenamefont {Wan}\ \emph {et~al.}(2018)\citenamefont {Wan},
  \citenamefont {Vlachas}, \citenamefont {Koumoutsakos},\ and\ \citenamefont
  {Sapsis}}]{Wan_2018}%
  \BibitemOpen
  \bibfield  {author} {\bibinfo {author} {\bibfnamefont {Z.~Y.}\ \bibnamefont
  {Wan}}, \bibinfo {author} {\bibfnamefont {P.}~\bibnamefont {Vlachas}},
  \bibinfo {author} {\bibfnamefont {P.}~\bibnamefont {Koumoutsakos}}, \ and\
  \bibinfo {author} {\bibfnamefont {T.}~\bibnamefont {Sapsis}},\ }\href@noop {}
  {\bibfield  {journal} {\bibinfo  {journal} {PloS one}\ }\textbf {\bibinfo
  {volume} {13}},\ \bibinfo {pages} {e0197704} (\bibinfo {year}
  {2018})}\BibitemShut {NoStop}%
\bibitem [{\citenamefont {Vlachas}\ \emph {et~al.}(2018)\citenamefont
  {Vlachas}, \citenamefont {Byeon}, \citenamefont {Wan}, \citenamefont
  {Sapsis},\ and\ \citenamefont {Koumoutsakos}}]{Vlachas_2018}%
  \BibitemOpen
  \bibfield  {author} {\bibinfo {author} {\bibfnamefont {P.~R.}\ \bibnamefont
  {Vlachas}}, \bibinfo {author} {\bibfnamefont {W.}~\bibnamefont {Byeon}},
  \bibinfo {author} {\bibfnamefont {Z.~Y.}\ \bibnamefont {Wan}}, \bibinfo
  {author} {\bibfnamefont {T.~P.}\ \bibnamefont {Sapsis}}, \ and\ \bibinfo
  {author} {\bibfnamefont {P.}~\bibnamefont {Koumoutsakos}},\ }\href@noop {}
  {\bibfield  {journal} {\bibinfo  {journal} {Proceedings of the Royal Society
  of London A: Mathematical, Physical and Engineering Sciences}\ }\textbf
  {\bibinfo {volume} {474}},\ \bibinfo {pages} {20170844} (\bibinfo {year}
  {2018})}\BibitemShut {NoStop}%
\bibitem [{\citenamefont {Verstraeten}\ \emph {et~al.}(2007)\citenamefont
  {Verstraeten}, \citenamefont {Schrauwen}, \citenamefont {D'Haene},\ and\
  \citenamefont {Stroobandt}}]{Verstraeten_2007}%
  \BibitemOpen
  \bibfield  {author} {\bibinfo {author} {\bibfnamefont {D.}~\bibnamefont
  {Verstraeten}}, \bibinfo {author} {\bibfnamefont {B.}~\bibnamefont
  {Schrauwen}}, \bibinfo {author} {\bibfnamefont {M.}~\bibnamefont {D'Haene}},
  \ and\ \bibinfo {author} {\bibfnamefont {D.~A.}\ \bibnamefont {Stroobandt}},\
  }\href@noop {} {\bibfield  {journal} {\bibinfo  {journal} {Neural Network}\
  }\textbf {\bibinfo {volume} {20}},\ \bibinfo {pages} {391} (\bibinfo {year}
  {2007})}\BibitemShut {NoStop}%
\bibitem [{\citenamefont {Inubushi}\ and\ \citenamefont
  {Yoshimura}(2017)}]{Inubushi_2017}%
  \BibitemOpen
  \bibfield  {author} {\bibinfo {author} {\bibfnamefont {M.}~\bibnamefont
  {Inubushi}}\ and\ \bibinfo {author} {\bibfnamefont {K.}~\bibnamefont
  {Yoshimura}},\ }\href@noop {} {\bibfield  {journal} {\bibinfo  {journal}
  {Scientific Reports}\ }\textbf {\bibinfo {volume} {7}},\ \bibinfo {pages}
  {10199} (\bibinfo {year} {2017})}\BibitemShut {NoStop}%
\bibitem [{\citenamefont {Lu}\ \emph {et~al.}(2017)\citenamefont {Lu},
  \citenamefont {Pathak}, \citenamefont {Hunt}, \citenamefont {Girvan},
  \citenamefont {Brockett},\ and\ \citenamefont {Ott}}]{Zhixin_2017}%
  \BibitemOpen
  \bibfield  {author} {\bibinfo {author} {\bibfnamefont {Z.}~\bibnamefont
  {Lu}}, \bibinfo {author} {\bibfnamefont {J.}~\bibnamefont {Pathak}}, \bibinfo
  {author} {\bibfnamefont {B.}~\bibnamefont {Hunt}}, \bibinfo {author}
  {\bibfnamefont {M.}~\bibnamefont {Girvan}}, \bibinfo {author} {\bibfnamefont
  {R.}~\bibnamefont {Brockett}}, \ and\ \bibinfo {author} {\bibfnamefont
  {E.}~\bibnamefont {Ott}},\ }\href@noop {} {\bibfield  {journal} {\bibinfo
  {journal} {Chaos}\ }\textbf {\bibinfo {volume} {27}},\ \bibinfo {pages}
  {041102} (\bibinfo {year} {2017})}\BibitemShut {NoStop}%
\bibitem [{\citenamefont {Pathak}\ \emph {et~al.}(2017)\citenamefont {Pathak},
  \citenamefont {Lu}, \citenamefont {Hunt}, \citenamefont {Girvan},\ and\
  \citenamefont {Ott}}]{Pathak_2017}%
  \BibitemOpen
  \bibfield  {author} {\bibinfo {author} {\bibfnamefont {J.}~\bibnamefont
  {Pathak}}, \bibinfo {author} {\bibfnamefont {Z.}~\bibnamefont {Lu}}, \bibinfo
  {author} {\bibfnamefont {B.}~\bibnamefont {Hunt}}, \bibinfo {author}
  {\bibfnamefont {M.}~\bibnamefont {Girvan}}, \ and\ \bibinfo {author}
  {\bibfnamefont {E.}~\bibnamefont {Ott}},\ }\href@noop {} {\bibfield
  {journal} {\bibinfo  {journal} {Chaos}\ }\textbf {\bibinfo {volume} {27}},\
  \bibinfo {pages} {121102} (\bibinfo {year} {2017})}\BibitemShut {NoStop}%
\bibitem [{\citenamefont {Ib{\'a}{\~n}ez-Soria}\ \emph
  {et~al.}(2018)\citenamefont {Ib{\'a}{\~n}ez-Soria}, \citenamefont
  {Garcia-Ojalvo}, \citenamefont {Soria-Frisch},\ and\ \citenamefont
  {Ruffini}}]{Ibanez_2018}%
  \BibitemOpen
  \bibfield  {author} {\bibinfo {author} {\bibfnamefont {D.}~\bibnamefont
  {Ib{\'a}{\~n}ez-Soria}}, \bibinfo {author} {\bibfnamefont {J.}~\bibnamefont
  {Garcia-Ojalvo}}, \bibinfo {author} {\bibfnamefont {A.}~\bibnamefont
  {Soria-Frisch}}, \ and\ \bibinfo {author} {\bibfnamefont {G.}~\bibnamefont
  {Ruffini}},\ }\href@noop {} {\bibfield  {journal} {\bibinfo  {journal}
  {Chaos}\ }\textbf {\bibinfo {volume} {28}},\ \bibinfo {pages} {033118}
  (\bibinfo {year} {2018})}\BibitemShut {NoStop}%
\bibitem [{\citenamefont {Pathak}\ \emph {et~al.}(2018)\citenamefont {Pathak},
  \citenamefont {Hunt}, \citenamefont {Girvan}, \citenamefont {Lu},\ and\
  \citenamefont {Ott}}]{Pathak_2018}%
  \BibitemOpen
  \bibfield  {author} {\bibinfo {author} {\bibfnamefont {J.}~\bibnamefont
  {Pathak}}, \bibinfo {author} {\bibfnamefont {B.}~\bibnamefont {Hunt}},
  \bibinfo {author} {\bibfnamefont {M.}~\bibnamefont {Girvan}}, \bibinfo
  {author} {\bibfnamefont {Z.}~\bibnamefont {Lu}}, \ and\ \bibinfo {author}
  {\bibfnamefont {E.}~\bibnamefont {Ott}},\ }\href@noop {} {\bibfield
  {journal} {\bibinfo  {journal} {Phys. Rev. Lett.}\ }\textbf {\bibinfo
  {volume} {120}},\ \bibinfo {pages} {024102} (\bibinfo {year}
  {2018})}\BibitemShut {NoStop}%
\bibitem [{\citenamefont {Antonik}\ \emph {et~al.}(2018)\citenamefont
  {Antonik}, \citenamefont {Gulina}, \citenamefont {Pauwels},\ and\
  \citenamefont {Massar}}]{Antonik_2018}%
  \BibitemOpen
  \bibfield  {author} {\bibinfo {author} {\bibfnamefont {P.}~\bibnamefont
  {Antonik}}, \bibinfo {author} {\bibfnamefont {M.}~\bibnamefont {Gulina}},
  \bibinfo {author} {\bibfnamefont {J.}~\bibnamefont {Pauwels}}, \ and\
  \bibinfo {author} {\bibfnamefont {S.}~\bibnamefont {Massar}},\ }\href
  {\doibase 10.1103/PhysRevE.98.012215} {\bibfield  {journal} {\bibinfo
  {journal} {Phys. Rev. E}\ }\textbf {\bibinfo {volume} {98}},\ \bibinfo
  {pages} {012215} (\bibinfo {year} {2018})}\BibitemShut {NoStop}%
\bibitem [{\citenamefont {Jaeger}(2001)}]{Jaeger_2001}%
  \BibitemOpen
  \bibfield  {author} {\bibinfo {author} {\bibfnamefont {H.}~\bibnamefont
  {Jaeger}},\ }\href@noop {} {\bibfield  {journal} {\bibinfo  {journal} {GMD
  Report}\ }\textbf {\bibinfo {volume} {148}},\ \bibinfo {pages} {13} (\bibinfo
  {year} {2001})}\BibitemShut {NoStop}%
\bibitem [{\citenamefont {Jaeger}\ and\ \citenamefont
  {Haas}(2004)}]{Jaeger_2004}%
  \BibitemOpen
  \bibfield  {author} {\bibinfo {author} {\bibfnamefont {H.}~\bibnamefont
  {Jaeger}}\ and\ \bibinfo {author} {\bibfnamefont {H.}~\bibnamefont {Haas}},\
  }\href@noop {} {\bibfield  {journal} {\bibinfo  {journal} {Scince}\ }\textbf
  {\bibinfo {volume} {304}},\ \bibinfo {pages} {78} (\bibinfo {year}
  {2004})}\BibitemShut {NoStop}%
\bibitem [{\citenamefont {Maass}\ \emph {et~al.}(2002)\citenamefont {Maass},
  \citenamefont {Natschl\"ager},\ and\ \citenamefont {Markram}}]{Maass_2002}%
  \BibitemOpen
  \bibfield  {author} {\bibinfo {author} {\bibfnamefont {W.}~\bibnamefont
  {Maass}}, \bibinfo {author} {\bibfnamefont {T.}~\bibnamefont
  {Natschl\"ager}}, \ and\ \bibinfo {author} {\bibfnamefont {H.}~\bibnamefont
  {Markram}},\ }\href@noop {} {\bibfield  {journal} {\bibinfo  {journal}
  {Neural Computation}\ }\textbf {\bibinfo {volume} {14}},\ \bibinfo {pages}
  {2531} (\bibinfo {year} {2002})}\BibitemShut {NoStop}%
\bibitem [{\citenamefont {Lu}\ \emph {et~al.}(2018)\citenamefont {Lu},
  \citenamefont {Hunt},\ and\ \citenamefont {Ott}}]{Lu_2018}%
  \BibitemOpen
  \bibfield  {author} {\bibinfo {author} {\bibfnamefont {Z.}~\bibnamefont
  {Lu}}, \bibinfo {author} {\bibfnamefont {B.~R.}\ \bibnamefont {Hunt}}, \ and\
  \bibinfo {author} {\bibfnamefont {E.}~\bibnamefont {Ott}},\ }\href {\doibase
  10.1063/1.5039508} {\bibfield  {journal} {\bibinfo  {journal} {Chaos}\
  }\textbf {\bibinfo {volume} {28}},\ \bibinfo {pages} {061104} (\bibinfo
  {year} {2018})},\ \Eprint
  {http://arxiv.org/abs/https://doi.org/10.1063/1.5039508}
  {https://doi.org/10.1063/1.5039508} \BibitemShut {NoStop}%
\bibitem [{\citenamefont {Lukosevivcius}\ and\ \citenamefont
  {Jaeger}(2009)}]{Lukosevicius_2009}%
  \BibitemOpen
  \bibfield  {author} {\bibinfo {author} {\bibfnamefont {M.}~\bibnamefont
  {Lukosevivcius}}\ and\ \bibinfo {author} {\bibfnamefont {H.}~\bibnamefont
  {Jaeger}},\ }\href@noop {} {\bibfield  {journal} {\bibinfo  {journal}
  {Computer Science Review}\ }\textbf {\bibinfo {volume} {3}},\ \bibinfo
  {pages} {127} (\bibinfo {year} {2009})}\BibitemShut {NoStop}%
\bibitem [{\citenamefont {Ishioka}(1999)}]{ishioka_1999}%
  \BibitemOpen
  \bibfield  {author} {\bibinfo {author} {\bibfnamefont {K.}~\bibnamefont
  {Ishioka}},\ }\href@noop {} {\enquote {\bibinfo {title} {ispack-0.4.1},}\
  }\bibinfo {howpublished} {\url{http://www.gfd-dennou.org/arch/ispack/, }}
  (\bibinfo {year} {1999}),\ \bibinfo {note} {{G}FD Dennou Club}\BibitemShut
  {NoStop}%
\bibitem [{\citenamefont {Takens}(1981)}]{Takens}%
  \BibitemOpen
  \bibfield  {author} {\bibinfo {author} {\bibfnamefont {F.}~\bibnamefont
  {Takens}},\ }in\ \href@noop {} {\emph {\bibinfo {booktitle} {Dynamical
  systems and turbulence, {W}arwick 1980 ({C}oventry, 1979/1980)}}},\ \bibinfo
  {series} {Lecture Notes in Math.}, Vol.\ \bibinfo {volume} {898}\ (\bibinfo
  {publisher} {Springer, Berlin-New York},\ \bibinfo {year} {1981})\ pp.\
  \bibinfo {pages} {366--381}\BibitemShut {NoStop}%
\bibitem [{\citenamefont {Sauer}\ \emph {et~al.}(1991)\citenamefont {Sauer},
  \citenamefont {Yorke},\ and\ \citenamefont {Casdagli}}]{Embedology}%
  \BibitemOpen
  \bibfield  {author} {\bibinfo {author} {\bibfnamefont {T.}~\bibnamefont
  {Sauer}}, \bibinfo {author} {\bibfnamefont {J.~A.}\ \bibnamefont {Yorke}}, \
  and\ \bibinfo {author} {\bibfnamefont {M.}~\bibnamefont {Casdagli}},\
  }\href@noop {} {\bibfield  {journal} {\bibinfo  {journal} {J. Stat. Phys.}\
  }\textbf {\bibinfo {volume} {65}},\ \bibinfo {pages} {579} (\bibinfo {year}
  {1991})}\BibitemShut {NoStop}%
\end{thebibliography}%

\end{document}